\documentclass{aastex6}

\usepackage{graphicx}
\usepackage{epstopdf}

\shorttitle{Gravitational Microlensing Events as a Target for SETI project}
\shortauthors{Rahvar}

\begin{document}
\title{Gravitational Microlensing Events as a Target for SETI project}

\author{Sohrab Rahvar\altaffilmark{1}}
\affil{Department of Physics, Sharif University of
Technology, P.O.Box 11365--9161, Tehran, Iran}
\email{rahvar@sharif.edu}

\begin{abstract}
Detection of signals from a possible extrasolar technological
civilization is one of the challenging efforts of science. In this work, we propose using natural telescopes made of single or binary gravitational lensing systems to magnify leakage of electromagnetic signals from a remote planet harbours an Extra Terrestrial Intelligent (ETI) technology. The gravitational microlensing surveys are monitoring a large area of Galactic
bulge for searching microlensing events and they find more than $2000$ 
events per year. These lenses are capable of playing the role of natural telescopes and in some occasions they can magnify radio band signals from the planets orbiting around the source stars in gravitational microlensing systems. Assuming that frequency of electromagnetic waves used for telecommunication in ETIs is similar to ours, we propose follow-up observation of microlensing events with radio telescopes such as  Square Kilometre Array (SKA), Low Frequency Demonstrators (LFD) and Mileura Wide-Field Array (MWA). Amplifying signals from the leakage of broadcasting by an Earth-like civilizations will allow us to detect them up to center of Milky Way galaxy.  Our analysis shows that in binary microlensing systems, the probability of amplification of signals from ETIs is more than that in single microlensing events. Finally we propose target of opportunity mode for 
 follow-up observations of binary microlensing events with  SKA as a new observational program for searching ETIs. 
Using the optimistic values for the factors of Drake equation 
provides detection of about one event per year. \\
\end{abstract}

\section{Introduction}
 
Are we alone in the Universe ? This is one of the deepest questions of
mankind and in recent years there have been a lot of efforts to answer
this question. The Search for Extra Terrestrial Intelligence (SETI) is one of the projects
trying to answer this question and since 1960s  no confirmed 
signal from extraterrestrial life has been detected~\citep{wilson}.  Another important effort started with direct and indirect observations of
extra-solar planets after the discovery of first exoplanet by  
\cite{mayer}. Nowadays, projects like Kepler \footnote{http://www.nasa.gov/kepler} increased
the number of extrasolar planets to more than one thousand candidates \citep{kepler}  and an
important question still is, are there any intelligent life exists in one of the discovered planets ?

In order to detect Extra Terrestrial Intelligent (ETI), we need to listen  leakages from the telecommunication signals of advanced technology civilizations. The transmitting signals can be either particles or photons. The properties of transmitting particle has been discussed extensively in \citet{cyc} with the following properties of (a) having minimum energy per quantum for transmitting particle,  (b)  large 
velocity for particle (c) particles easily generate and focused and finally (d) particles have not easily 
absorbed by the interstellar medium. The best candidate that satisfies these four properties is the electromagnetic waves in radio or microwave wavelengths . The main challenge of detection of signals from an advanced civilization is that in the large
astronomical scales signals are faded out. Also, in the radio waves for frequencies smaller than  $100$MHz (larger than $3$ meter), the free-free absorption become important and for frequencies larger than $1$THz (smaller than 300 micro-meter) absorption becomes increasingly important.

Assuming that the electromagnetic signals are transmitted by an advance civilization, these signals should have the specific property of violating the natural transmission mechanism to be distinguished from the natural sources \citep{Tarter}. Moreover, we expect that transmission by the 
intelligent life has to be done in narrow band by
the artificial sources with the minimum amount of $\Delta\lambda/\lambda$, similar to what 
our civilization is doing for transmitting signals. With our primitive technology, we can produce $\Delta \lambda/\lambda \sim 10^{-12}$  compare to nature that is around $10^{-4}$ to $10^{-3}$ in astrophysical systems and water masers with limiting value of $10^{-6}$. Assuming that Earth-like advanced civilizations using the same range of electromagnetic waves for telecommunications that we use, the present radio telescopes can eavesdrop their radio transmissions \citep{loeb}.

The aim of this work is using gravitational microlensing as natural amplifiers to magnify leakage of signals from remote planets at the Galactic scales. The gravitational lenses can magnify
electromagnetic waves from distant sources by converging light beams from the source star. This light magnification can be done either by single or 
binary lenses. At the present time, there are two main observational campaigns of OGLE and 
MOA that use  wide field telecopies to monitor centre of Galaxy for gravitational microlensing detections. In 2014, OGLE could detect over 2000 microlensing events \citep{moa,gaudi}. The next generation of microlensing surveys as Korean Microlensing Telescope Network 
(KMTNet) is going to cover a larger area and discover more microlensing events than the present experiments \citep{KMTNet}.


The main propose of this work is the follow-up observation of microlensing events with radio telescopes to detect transmitting signals from planets hosting ETI and orbiting around the source stars in the microlensing systems. During the gravitational microlensing, electromagnetic waves independent of their wave-lengths are magnified from both 
the parent source star and a planet orbiting around it. However,  since the source star and the associated planet are separated with the distance in the order of astronomical unit, their magnifications do not peaks at the same time. This method increases the depth of SETI observation from few parsecs up to the centre of Galaxy.  This technique in the visible band also has been proposed for detecting hot Jupiters by illuminating planets around the parent stars \citep{rahvarsajadian}. 


The organization of this paper is as follows: In section (\ref{rsources}), we introduce the radio sources on an Earth-like planet  similar to our transmitters on the Earth and study the possibility of detecting these sources with the present instruments that are using in the cosmological observations. In section (\ref{introduction}), we introduce the basic formalism of gravitational microlensing as a natural amplifier of signals from an Earth-like planet by single and binary lensing systems. In Section (\ref{statistics}), we perform a Monte-Carlo simulation and estimate the number of high magnification microlensing events of planets by follow-up radio observations for both single and binary lensing systems.  In this section we also investigate the wave optics effects of gravitational microlensing in our observation and the suppression of very high magnified singles in the radio light curve. Details about this effect is given in appendix (\ref{A}). In Section (\ref{statint}) after introducing Drake equation, we investigate the number of expected candidates with signals from the planets with intelligent life. The conclusion is given in section (\ref{conc}).

\section{Radio sources from an Earth-like civilization}
\label{rsources}
\begin{table*}
\begin{center}
\begin{tabular}{|c|c|c|c|c|c|c|}
\hline\hline
Service  & Freq.   & Transmitters & Max. Power  &  Bandwidth   & Power  & Power/Hz \\ 
 (1) & (2)~(MHz) &(3)~(No.) & (4)~ per Tr. (W) &(5)~ (Hz) & (6)~(W) & (7)~(W/Hz) \\ \hline
Military & $\sim 400$ & 10 & $2 \times 10^{11}$ & $10^{3}$ & $2 \times 10^{12}$ & $2 \times 10^9$\\
FM & 88-108 & 9000 & $4 \times 10^{3}$ & 0.1 & $4 \times 10^{7}$ & $4 \times 10^{8}$  \\
TV & 40-850 & 2000 & $5 \times 10^{5}$ & 0.1 & $10^{9}$ & $10^{10}$\\
\hline \hline
\end{tabular}
\caption{Characteristics of the strongest radio transmitters on the Earth, adopted from \cite{sullivan} and \cite{loeb} .The power of military transmitter is corrected based on \cite{loeb}. The first column is the type of service, second column is the frequency range of transmission, third column is the number of transmitters, forth column is the maximum power per transmitter, fifth column is the bandwidth, sixth column is the total power of transmitters and the seventh column is the transmitter power per Hertz. Moreover, Arecibo telescope has four radar that transmitting with the power of $2\times 10^{13}$ W \citep{Tarter}.}
\label{table1}
\end{center}
\end{table*}

Let us imaging an Earth-Like civilization that is producing at the same frequency rang and the same power of electromagnetic waves that we are producing on the Earth. We note that according to the classification of advanced civilizations by Kardashev, these planets may produce stronger transmissions compare to the Earth \citep{kardashev} or due to improving fibre optics communication on their planet, they produce less man-made noises compare to ours. The transmitted signals 
might be either in beacon mode or unintentional leakage of signals. Here we use the later mode of transmission in our study \citep{loeb}. As we discussed before, the artificial signals are narrow band and encoded signals contains more information compare to the natural signals. The complexity of signals from intelligent civilization can be measured by various mathematical methods.

Different types of radio sources as a target for intelligent civilization has been discussed in \cite{sullivan}. According to Table (\ref{table1}) which is adapted from \cite{sullivan} and \cite{loeb}, military radars with maximum power per transmitter ($P_{max} \simeq 2\times 10^{11}$~W) is the strongest sources that radiate radio-waves isotropically. Television and FM stations are the second and third ranks in the power of radiation.  Moreover, Arecibo Observatory uses transmitters that use them occasionally for ETI communications. This observatory has four radar transmitters, with effective isotropic radiated powers of $20$ TW at $2380$ MHz, $2.5$ TW (pulse peak) at $430$ MHz, $300$ MW at $47$ MHz, and $6$ MW at 8 MHz \citep{Tarter}. Assuming an Earth-like planet in our Galaxy, SETI project can investigate the following characteristics to search for the signals of intelligent life, (a) short bandwidth transmission, (b) frequency and intensity modulation due to rotation of planet around its axis and around the parent star and (c) measurement of complexity in data encoded in the signals.

In recent years the cosmological experiments are designed to observe $21$cm emissions of HII regions from redshift of $z\simeq 6-12$, corresponds to $200-400$~MHz \citep{reion}. These cosmological observations are aiming to observe the reionization history of Universe. Fortunately, this range of frequency overlaps with the frequency range of telecommunications on the Earth (see Table \ref{table1}), and it can be used for observation of Earth-like planets in our Galaxy (if intelligent life uses the same range of frequency). While military transmitters are the strongest radio sources, TV and FM radio sources with $10-100$~kW can outshine Sun up to almost $100$ times, means that radio emission from parent star does not blend with signals from an Earth-like planet. So the main limitation on the observation is the threshold of signal to noise ratio of receivers that we are using on the Earth. The signal to noise increases by increasing the square root of exposure time of observation, however exposures longer than the spinning period of planet or orbital period of planet around the parent star causes broadening of emission signals and erasing the modulation effect. On the other hand for long exposures, we can not detect complex signals from the radio flux by accumulating signals. Here we 
study the sensitivity of MWA-LFD and SKA for detection of signals from the intelligent life. 



\begin{figure}
\begin{center}
\includegraphics[width=80mm]{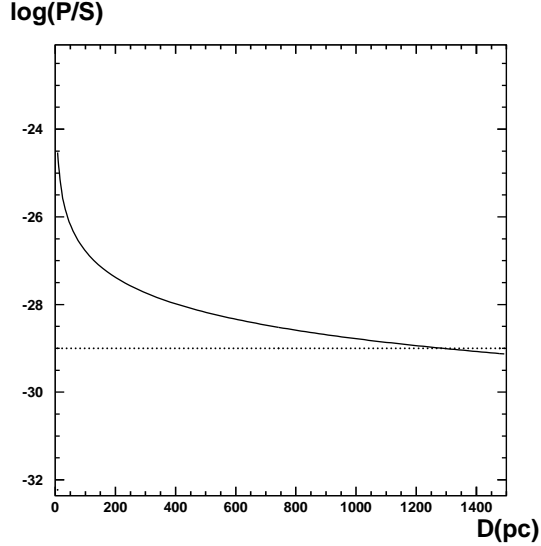}
\end{center}
\caption{\label{limit}
 The logarithm of Power per area (in $W/m^2$) from a Military-like source  adapted from Table (\ref{table1}) transmitted from an 
Earth-like intelligent life as a function of distance from the source (in parsec). The horizontal dashed line (i.e. $P/S = 10^{-29} W/m^2$) is the limiting value that SKA can observe signals, means that the source is detectable up to distance of $D \simeq 1.26$ kpc.}
\end{figure}

For the case of MWA-LFD, the point source sensitivity (PPS) at $200$MHz is \footnote{The value of FPSS for other frequencies can be found at http://web.haystack.mit.edu/arrays/MWA/LFD}
\begin{equation}
F_{PSS} = 0.4 ~ mJy~ \left(\frac{\Delta\nu}{8 { kHz}}\frac{t_{obs}}{{ month}}\right)^{-1/2},
\label{mwa}
\end{equation}
where $t_{obs}$ is the total exposure time and $\Delta\nu$ is the bandwidth. For a source with a minimum power of $P_{min}$ located at the distance of $D$ and emitting with the bandwidth of $\Delta\nu$, the threshold of observability from equation (\ref{mwa}) is
\begin{equation}
P_{min} = 1.99\times 10^{13} {W} \times \left(\frac{D}{100 ~pc}\right)^2\times\left(\frac{\Delta\nu}{8~kHz}\right)^{1/2}\times \left(\frac{t_{obs}}{10~{ min}}\right)^{-1/2}.
\end{equation}
Let us assume that we perform observations with the exposure time of $10$~min. This strategy of observation enable us to monitor modulation of frequency within few hours from the spin of an Earth like planet. Comparing the minimum flux of detection with the power of military transmitters in Table (\ref{table1}) as well as Arecibo transmitter, we can detect them up to $10$pc and within this volume there are almost thousand stars available. This volume can be increased with increasing the exposure time of observation, however we will miss the sensitivity to the modulation.

The other important telescope in our desired range of frequency is the Square
Kilo-meter Array (SKA) telescope. This telescope also is designed for astrophysical and cosmological studies. It has the threshold sensitivity of $10^{-28}~W/m^2$ and is able to monitor sources with high cadence. This will enable us to monitor time variation of source flux and search for possible meaningful signals from an Earth-like planet.  By using narrow bandwidth of detection, it is possible to improve the threshold sensitivity to  $10^{-29}~W/m^2$ \citep{ska}. Using the strongest transmitter on an Earth-like planet from Table \ref{table1}, we can put limit on distance that can be detected with SKA. The result from Figure (\ref{limit}) is $D_0 \simeq1.26$ kpc.

In the following sections we will introduce gravitational microlensing as the natural amplifiers of signals from remote planets. We will provide an observational strategy based on 
observational mode of microlensing surveys and estimate the number of planets that are illuminated with the gravitational microlensing in radio band.

\section{Gravitational Lensing as Natural Telescopes}
\label{introduction}
 The gravitational lensing of a star by another star in the Milky Way galaxy, for the first time is introduced by Einstein in (1936) and since angular resolution of an optical telescope is less than the Einstein ring,  Einstein noted in his paper that " {\it  There is no great chance of observing this phenomenon}" \cite{Einstein}. On the other hand the probability of detecting gravitational lensing inside the Galaxy is of the order of $10^{-7}$.  After five decades, Paczy\'nski in 1986 revisited the question of gravitational microlensing and estimate the observability of this phenomenon inside the Galaxy and propose the application of it for searching 
compact dark matter objects inside the Galactic halo \cite{pac}. 

In the gravitational lensing inside the Milky Way, while the angular separation between the images is small and by the ground-based telescopes images are unresolvable, the magnification of light from the source star during the transit of lens is a measurable quantity. The magnification depends on the angular separation between the source and lens stars. The overall magnification from single lensing is given by 
\begin{equation}
A(t) = \frac{2 +u(t)^2}{u(t)\sqrt{4 + u(t)^2}},
\label{pac}
\end{equation}
where $u$ is the impact parameter normalized to the Einstein ring. The impact parameter is a dynamical 
parameter and for a single lens changes by time as $u(t)^2 = u_0^2 +(t - t_0)^2/ t_E^{2} $ where 
$u_0$ is the closest distance from the lens, $t_0$ is the moment of maximum magnification and $t_E$ is the Einstein crossing time \citep{pac,rahvar}. 

\begin{figure}
\begin{center}
\includegraphics[width=80mm]{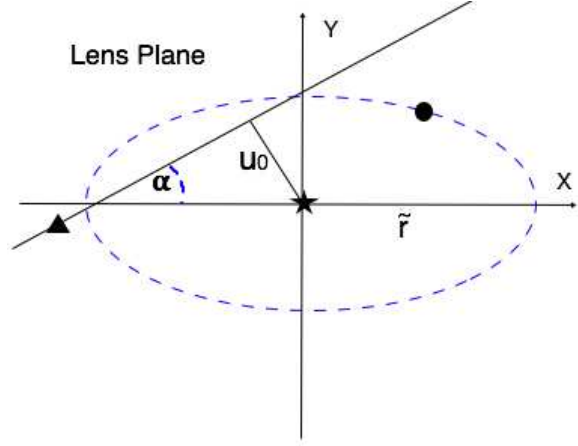}
\end{center}
\caption{\label{lensplanet}  The projection of source star (represented by star) and the 
orbital path of planet (elliptical) on the lens plane. The lens (triangle) moves 
on a straight line on the lens plane. The minimum distance of lens to the source star is given by $u_0$ and $\tilde{r}$ is the projected orbital radius of planet on the lens plane, normalized to the Einstein radius.}
\end{figure}

We imagine that in addition to the source star, there is a planet with intelligent life orbiting around the source star. The projected orbit of planet on the lens plane and relative distance of these three objects is shown in Figure (\ref{lensplanet}). 
We take $\tilde{r}$ as the orbital radius of planet projected on the 
lens plane, normalized to the Einstein radius of lens (i.e. $\tilde{r} = \frac{D_{ol}}{D_{os}}\frac{r}{R_E}$), $\omega$ as the angular velocity of planet and  $\phi$ as the initial phase of planet.  The equation of motion of planet in this coordinate system is given by 
\begin{eqnarray}
X_p &=& \tilde{r} \cos(\omega t +\phi), \nonumber \\
Y_p &=& \tilde{r} \sin(\omega t +\phi) \sin\beta,
\label{sl}
\end{eqnarray}
where $90-\beta$ is the angle between the Y-axis and the orbital plane and the orbital plane goes through the X-axis.
The parent star is located at the centre of coordinate system and the trajectory of lens is given by 
\begin{eqnarray}
X_L &=& (\frac{t -t_0}{t_E})\cos\alpha - u_0\sin\alpha, \nonumber \\
Y_L &=& (\frac{t -t_0}{t_E})\sin\alpha + u_0\cos\alpha, 
\label{lens}
\end{eqnarray}


where $\alpha$ is the angle between the lens trajectory and semi-major axis and $u_0$ is the minimum impact parameter compare to the source star. The distance between the projected position of planet on the lens plane and lens, $u_{PL}$, is calculated based on relative distance (projected on the lens plane) between the two objects as follows:
\begin{eqnarray}
 u_{PL}^2(t; t_0,u_0,t_E,\alpha,\omega,\tilde{r},\beta) &=&   \left[(\frac{t -t_0}{t_E})\cos\alpha - u_0\sin\alpha -  \tilde{r} \cos(\omega t +\phi)\right]^2 \nonumber \\
 &+& \left[ (\frac{t -t_0}{t_E})\sin\alpha + u_0\cos\alpha - \tilde{r} \sin(\omega t +\phi) \sin\beta\right]^2.
\label{imp}
\end{eqnarray}
\begin{figure}
\begin{center}
\includegraphics[width=80mm]{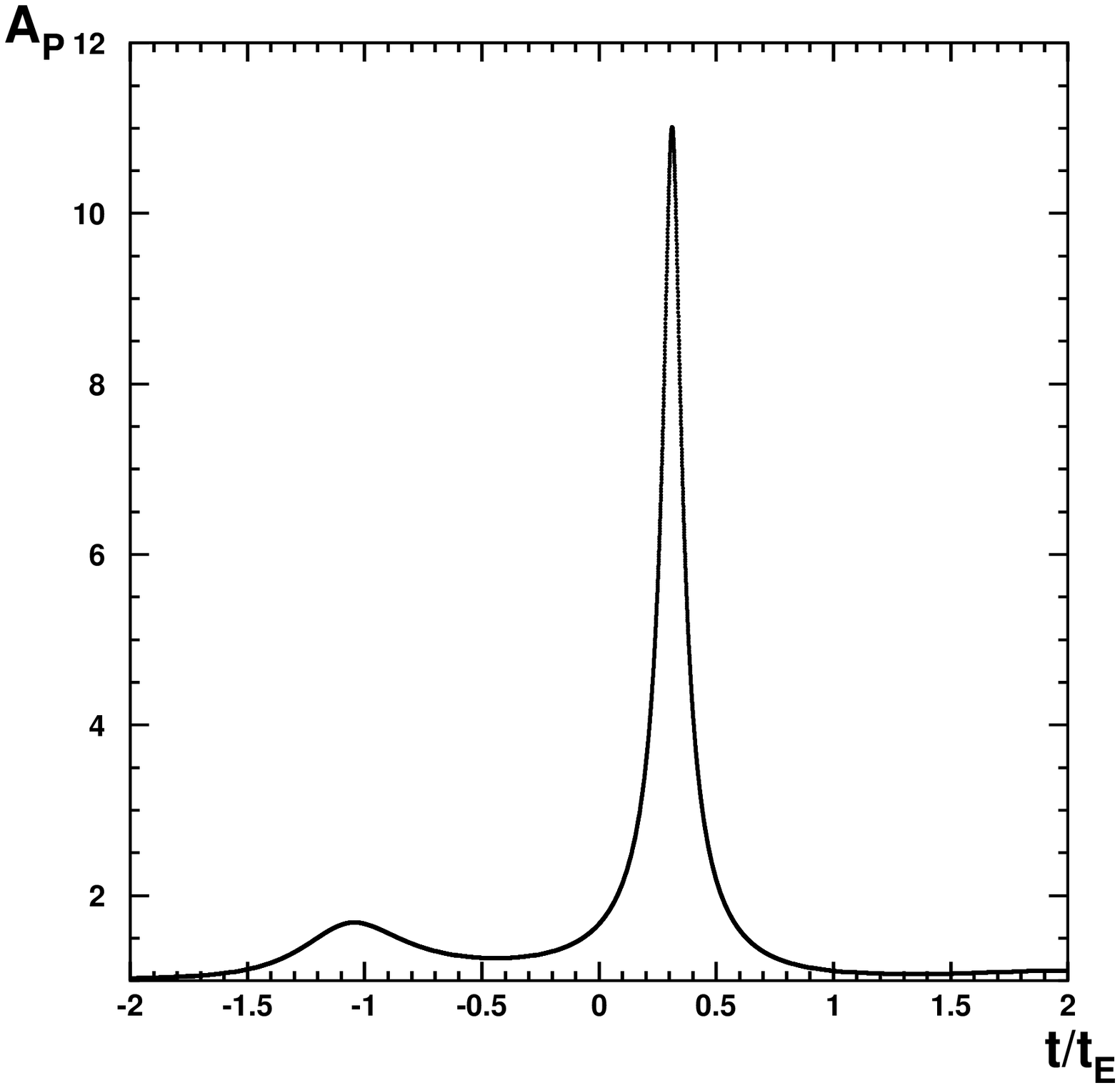}
\includegraphics[width=80mm]{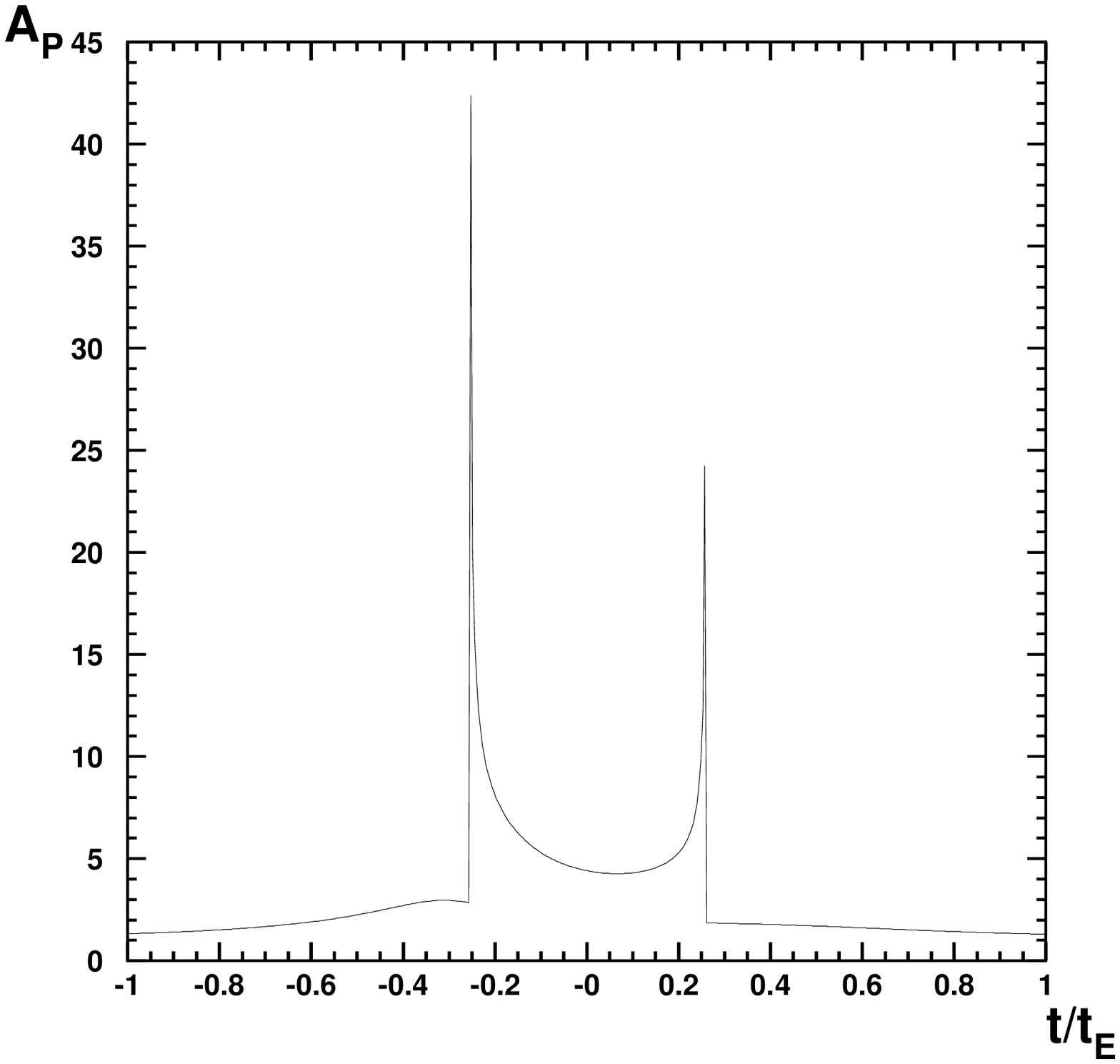}
\end{center}
\caption{\label{lc1} Typical magnification of flux from a planet as a function of time (normalized to the Einstein crossing time). Left panel represents microlensing light curve by a single lensing and right panel by a binary lensing. 
}
\end{figure}
We can calculate the magnification of planet's light by substituting this impact parameter in magnification formula for single lensing in equation (\ref{pac}). Here the impact parameter, $u_{PL}(t)$, depends on various orbital parameters of planet as $\tilde{r}$, $\omega$ and $\beta$ and the lens parameters as 
$t_E$, $u_0$ and $\alpha$.  

Figure (\ref{lc1}) on the left panel, represents the microlensing magnification factor of a planet orbiting around the parent star by single lensing. We note that the observation of microlensing events in short and long wavelengths has a main difference where for the long wavelengths and a point--like source, we have to take into account the wave optics features \citep{ms}. In this case images on the lens plane plays the role of slits as in the Young experiment and result is producing diffraction pattern on the observer position. The observational consequence is reducing the magnification factor of lensing and avoiding mathematical singularities in the geometric optics. This effect is discussed in Appendix  (\ref{A}).

In the case of binary lensing, the formalism of lensing is complicated than single lensing \citep{witt}. In this case binary lens produces caustic lines (or singular lines) on the source plane. Whenever source crosses these lines, the magnification gets infinite. However in reality, due to finite size of source star and interference of coherent light, the magnification is a finite value. Figure (\ref{lc1}), on the right panel shows a typical light curve of a binary lens magnifying signals from a planet orbiting a source star. Here we consider the geometric optics in calculation of magnification \citep{martin}. 

In the next section we estimate the number of planets that their signals from intelligent civilizations can 
be illuminated by single or binary lensing up to the observational threshold of SKA telescope.


\section{Statistics of planet illumination}

\label{statistics}
\begin{figure}
\begin{center}
\includegraphics[width=80mm]{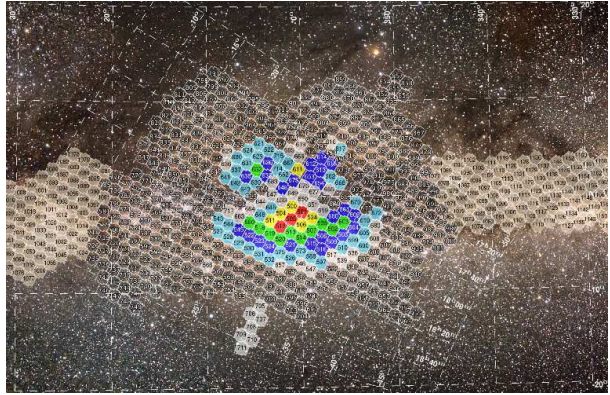}
\end{center}
\caption{\label{ogle} OGLE field for microlensing events in the direction of Galactic Bulge.  This figure is adapted from \cite{oglefield}.}
\end{figure}

In this section we estimate the number of Earth like planets that their signals can be amplified with gravitational microlensing (as a natural telescope).  The observation of microlensing events are done in two steps of (a) the survey mode, where survey groups are monitoring stars in the direction of Galactic Bulge and by ongoing analysis of light curves, alert the microlensing events.  (b) The follow-up mode, where telescopes around the globe observe microlensing candidates with high cadence and better photometric precision. The map of survey for OGLE experiment is shown in Figure (\ref{ogle}) towards the direction of Galactic Bulge . In this section, we will estimate the statistics of high amplification of Earth-like planets by single and binary lensing systems.


\subsection{Single Lens}

One of the crucial parameters in simulation of this effect in single and double lensing is determination of  $\tilde{r}$ as the projected orbit of planet on the lens plane, normalized to the Einstein radius as follows
\begin{equation}
\tilde{r}  = \frac{r_H(M_S)}{5.7~{AU}}\left(\frac{M_L}{0.5M_\odot}\right)^{-1/2}\left(\frac{D_S}{8 {kpc}}\right)^{-1/2}\left(\frac{x}{1-x}\right)^{1/2},
\label{rtidle}
\end{equation}
where $x = D_L/D_S$ represents the ratio of observer--lens to the observer--source distances, $M_L$ is the mass of lens and $r_H$ is the orbital radius of a planet assuming that it is located at the habitable zone of a source star. We note that $r_H$ is a function of mass and the age of source star that planet orbits around it. For the main sequence stars, it is given by $r_H (M_S) = (0.8-1.5)\times ({M_S}/{M_\odot})^2$ AU \citep{hs,hs2}.  For the red giants, while these type of stars are unstable, however for a solar type star stay at the first stages of its post main-sequence evolution, the temporal transit of the habitable zone is estimated to be of several $10^9$ years at $2$ AU and around $10^8$ years at $9$ AU \citep{gianthz}. 

\begin{figure}
\begin{center}
\includegraphics[width=80mm]{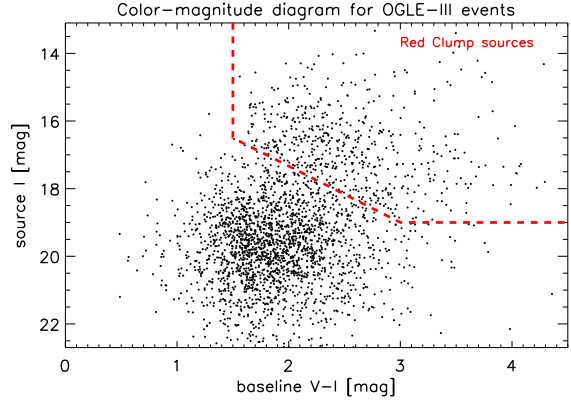}
\end{center}
\caption{\label{ocm} Color-magnitude of source stars of microlensing events in OGLE III survey from year 2001-2009 in the direction of Galactic bulge. The magnitude of source stars are unblended after imposing the blending factor from the best fit.  The top-right area of this diagram represents the red-clump source stars that comprises $\sim 16\%$ of all microlensing candidates.  This figure is adapted from \cite{oeffi}.}
\end{figure}
Fortunately from studying the color-magnitude distribution of source stars in OGLE III microlensing events, we can estimate the statistics of stars in the main-sequence and in the red giant area. For OGLE III in the direction of Galactic bulge, we adapt statistics of microlensing events from the observations in the period of years 2001-2009. Analyzing the distribution of stars in Figure (\ref{ocm}), the fraction of red-clump source stars to the overall number of stars is $16$ percent.  We use the same combination for the population of source stars in our Monte-Carlo simulation, associating  
a habitable zone to the source stars.



Taking into account that the limiting distance for detection of a military transmitter from an Earth-like planet with the SKA telescope is $D_0 = 1.26$~kpc (for details refer to Section \ref{rsources}), the minimum magnification factor from microlensing is related to detectability with SKA telescope at the distance of $D$ as follows
\begin{equation}
A_{min}> \left(\frac{D_S}{D_0}\right)^2.
\label{minimum}
\end{equation}
For the case of microlensing events in the direction of Galactic Bulge, if we assume for instance a source star located at $D_S\simeq 7$~kpc, the minimum magnification for the planet detection is $A_{min}>30$. For the microlensing events in the direction of Galactic Bulge, the source stars are not located at a fixed distance, rather they are distributed along our line of sight from the position of the observer to the centre of Galaxy. The main sequence stars are mainly belong to the disk and red clumps belong to the Galactic bulge.

In order to estimate the fraction of events with a magnification that satisfy the condition in expression (\ref{minimum}), we perform a Monte-Carlo simulation where the geometrical parameters of lens is chosen uniformly in the range of $u_0\in[0,1]$, $\alpha\in[0,2\pi]$, $\beta\in[0,\pi]$ and $\phi\in[0,2\pi]$. The other physical parameters as the Einstein crossing time $t_E$ and $\tilde{r}$ are taken according to the distribution of matter, dynamics of Galaxy and the mass function of lens stars in the Galaxy. The mass function including main sequence, brown dwarfs and compact stars are given with $\xi(\log m) = {dn}/{d \log m}$, as follows \cite{chabrier}
\begin{eqnarray}
\xi(\log m) &=& 0.093\times \exp\left[ - \frac{(\log m - \log 0,2)^2}{2\times (0.55)^2}   \right], ~~~ m\leq m_\odot  \nonumber \\
\xi(\log m) &=&  0.041\times m^{-1.35}, m>m_\odot
\end{eqnarray}
On the other hand, for simulating the source of the microlensing events, we adapt a complete sample from the Hipparcos catalog \cite{hipp} and enhance the contribution of red clumps to be comparable with the observation in the Galactic centre.

For the distribution of matter in the Galaxy, the density of lenses composed of density of disk and bulge is described with a thin disk and a central bar structure. The disk is exponential model adapted from Besancon model 
\cite{robin}
\begin{eqnarray}
\rho_{d}(r,z)  &=&  \rho_0\left[ \exp(-\frac{a^2}{h_{R_+}^2} ) -\exp(-\frac{a^2}{h_{R_-}^2} ) \right]~~~~~~~~~~~~~~~~~~~~~~~~~~~~~ age<0.15~Gyr, \\
\rho_{d}(r,z)  &=&  \rho_0\left[ \exp(- ( 0.5^2 + \frac{a^2}{h_{R_+}^2})^{1/2}) -\exp(-( 0.5^2 +\frac{a^2}{h_{R_-}^2} )^{1/2}) \right]~ age>0.15~Gyr,
\end{eqnarray}
where $h_{R_+} = 5.0$ kpc and  $h_{R_-} = 3.0$ for age$<0.15$ Gyr and $h_{R_+} = 2.53$ kpc and  $h_{R_-} = 1.32$ for age$>0.15$ Gyr,  $a^2 = r^2 + (z/\epsilon)^2$. $\epsilon$ and $\rho_0$ for different ages of stars are 
given in detail in \cite{robin}.


The bar is described in
a cartesian frame positioned at the Galactic center with the major
axis x tilted by $\Phi= 13$ degree with respect to the
Galactic center-Sun direction. The number density of starts in the bulge also is given by \cite{robin} 
\begin{eqnarray}
n_B(x,y,z)&=& n_0 e^{-r^2/2},~~~~~~~~~~~~~~~~~~~~~~~~~~ \sqrt{x^2+y^2}<R_c \\
n_B(x,y,z)&=& n_0 e^{-r^2/2} e^{-2(\sqrt{x^2+y^2} - R_c)^2},~~~~~ \sqrt{x^2+y^2}>R_c 
\end{eqnarray}

where $r^2 = \left[\left[(\frac{x}{a})^2 + (\frac{y}{b})^2 \right]^2 + (\frac{z}{c})^4\right]^{1/2}$, 
$n_0= 13.7 pc^{-3}$, $a=1.59$ kpc,
$b = 0.424$ kpc, $c=0.424$~kpc and $R_c=2.54$ kpc are the scale length factors.


 For the dynamics of Galaxy we apply separately the dynamics of stars in the disk and bulge. The global rotation of the disk is given as a function of the galactocentric distance by
\begin{equation}
v_{rot}(r) = v_{rot,\odot}\times 1.00762\left(\frac{r}{R_\odot} \right)^{0.0394} + 0.00712
\end{equation}
where $v_{rot,\odot} = 220 km/s$ \cite{brand}. The peculiar velocity of the disk stars is described by an anisotropic Gaussian distribution with the following radial,
tangential and perpendicular velocity dispersions of $\sigma_r = 34$ km/s, $\sigma_\theta = 28$ km/and $\sigma_z = 20$ km/s  \cite{rahal}. The distribution of transverse speed of stars in the bulge also is given by 
\begin{equation} 
f_T(v_\bot) = \frac{1}{\sigma_{bulge}^2}v_\bot \exp\left(- \frac{v_\bot^2}{2\sigma_{bulge}^2} \right),
\end{equation}
where $\sigma_{bulge} \simeq 110$ km/s.

The relative transverse speed of source and lens on the lens plane is given by ${\bf v}_\bot = x({\bf v}_S - {\bf v}_E)_\bot - ({\bf v}_L - {\bf v}_E)_\bot$ where all the vectors define in two dimension parallel to the lens plane and perpendicluar to the line of sight. The rate of events, taking into account that each lens spans a tube with the diameter of $2 R_E$ and the length of $v_\bot \times T_{obs}$ is given by 
\begin{equation}
d\Gamma \propto \sqrt{M_L x(1-x)}\rho_\star(x) \xi(\log M_L)v_\bot f_S(v_S) f_L(v_L) dx dv_S dv_L d\log M_L,
\end{equation}
where $\rho_\star(x)$ is the stellar density of Galaxy, $\xi$ is the mass function of stars in logarithmic scale, $f_S(v_S)$ and  $f_L(v_L)$ are the velocity distribution functions of source and lens in two dimension. The important  observable of microlensing events is the duration of event which is given by $t_E$ and detectability of an event directly is related to this parameter. In our simulation, we adapt the detection efficiency 
function  (i.e. $\epsilon(t_E)$) from the OGLE survey \cite{oeffi}. 

\begin{figure}
\begin{center}
\includegraphics[width=80mm]{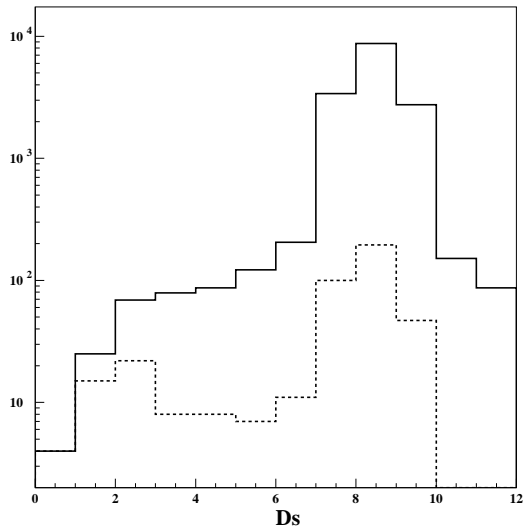}
\end{center}
\caption{\label{ds} The distribution of distance of source stars (i.e. $D_s$) for overall microlensing events from the Monte-Carlo simulation in solid line and the distribution of microlensing events satisfy the condition of (\ref{minimum}) with $2.5\%$ of overall events in the dashed line. The distribution is scaled in logarithmic scale.
Magnifying signals from planets in the microlensing event, we can detect them up to the centre of Galaxy}.
\end{figure}

For single-lens microlensing events in the direction of Galactic Bulge, assuming a planet orbiting at the habitable zone of source star, almost $2.5\%$ of overall observed microlensing events satisfies the condition of expression (\ref{minimum}). Figure (\ref{ds}) represents the distribution of microlensing events in term of distance of source stars (i.e. $D_s$) for overall microlensing (in solid line) and the distribution of microlensing events satisfy the condition of expression (\ref{minimum}) (in dashed line). The average distance of observer to the source stars for all the events is $8.36$ kpc and for those with positive signal from planet is $7.49$ kpc. While the observation of microlensing events with 
nearby sources improve receiving strong signals from the planets, estimation of distance of source star in the ongoing microlensing events is a challenging problem. 

To estimate the distance of source star,  the color-magnitude of stars around the source star within the  
radius of $60''$ is measured (as an example see \cite{dist}). Assuming that all the stars in the field are affected by the same amount of extinction, using the position of center of red clumps in the color-magnitude diagram from one hand and the position of red clump stars in the Hipparcos catalog \citep{pac3}, we can calculate the extinction of the field. Then the dereddened color and magnitude of the source star (i.e. $I_{s0},(V-I)_{s0}$) with measuring its offset with the red clump can be calculated. The result would be an estimation from the distance modulus of the source star as well as the angular size of the source star. Due to uncertainty in the distance measurement, we propose performing 
the radio follow-up observation for all the microlensing candidates regardless of their source distances.

In this part, we have introduced the gravitational microlensing as a natural amplifier to enhance signals from planets and enable us to detect them up to distances of $10$ Kpc. In another word, the depth of our observation with SKA could be extended up to the center of Galaxy. With the present technology, microlensing surveys observe almost $2000$ microlensing event per year. 
Regarding that the efficiency for planet illumination from single lensing in the Monte-Carlo simulation 
is about $2.5\%$, we would expect to detect almost $50$ high magnification of planets around the microlensing source stars. For these events the average value of orbital size of habitable planets normalized to the Einstein 
radius is $<\tilde{r}> = 0.18$,  compare to $<\tilde{r}> = 0.19$ which is the average value of this parameter 
for all the microlensing events regardless of planet detection. In the next section we will estimate the number of high magnification planets by the binary lensing. The probability of these planes hosting intelligent communicating life will be discussed later. 


\subsection{Binary lensing}
In the binary lensing, the lensing system composed of two objects. These objects can be either (a) a binary star or (b) a star with its companion planet or (c) any other combination of sufficiently high mass objects, such as black holes, brown dwarfs, etc. Increasing the number of lenses from one to two or more objects increases dramatically the complication of the lensing equation. One of the features of having multiple lensing is the caustic lines
on the source plane. Once a source crosses these lines, in theory we have infinite magnification and in the 
observation, we can get sparks of light from the source object. In the case of binary lensing, either source star or planet orbiting around it or both objects can cross the caustic lines. The advantage of using binary lensing for detecting planet is that with an early warning system, from the deviation of light curve from the single lensing, we can identify an ongoing binary microlensing event before the light curve peaks. We will show that the probability of
high magnification of planet in binary lensing is more than a typical single lens.

Investigating the early warning database of OLGE III surveys during years 2002-2003 results in detection of $15$  binary microlensing events out of overall $851$ microlensing candidates.  In another word, $1.76 \%$ of microlensing events in this catalog are binary lenses. \cite{oglebinary} found the best $\chi^2$ fit to the light curve and could reconstruct the caustic lines on the source plane or the projection of these lines on the lens plane.

\begin{figure}
\begin{center}
\includegraphics[width=80mm]{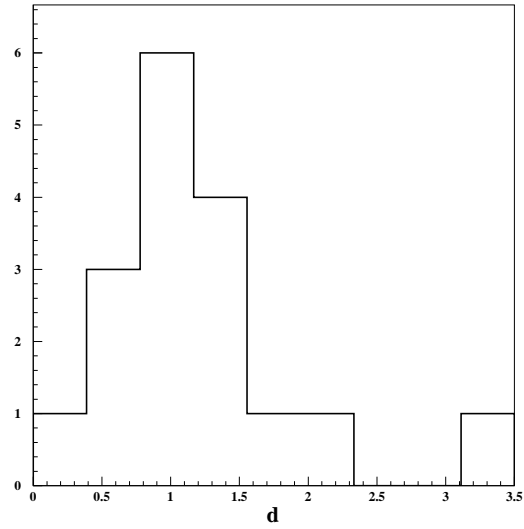}
\end{center}
\caption{\label{d} Distribution of distance between lenses of $15$ binary micolensing events from OGLE 2002-2003 microlensing list \citep{oglebinary}. The distance between the lenses are normalized to the Einstein radius of overall mass of lenses.}
\end{figure}


 In what follows we adapt the caustic structure of binary lensing from the best fit. The stellar type of source stars are estimated from the position of stars in the color-magnitude diagram and for each star we can calculate from equation (\ref{rtidle}) a habitable zone for the planet orbiting around the source star. Then we calculate the probability of crossing of a planet orbiting around the source star by the caustic lines of a binary lens. We adapt the distribution of distance between the lens stars as in the observation sample that is shown in Figure (\ref{d}).  For comparing the size of orbital radius of planets in the habitable zone with the caustic lines and probability of caustic crossing, we repet the Monte-Carlo simulation as in the single lensing. 
 

\begin{figure}
\begin{center}
\includegraphics[width=100mm]{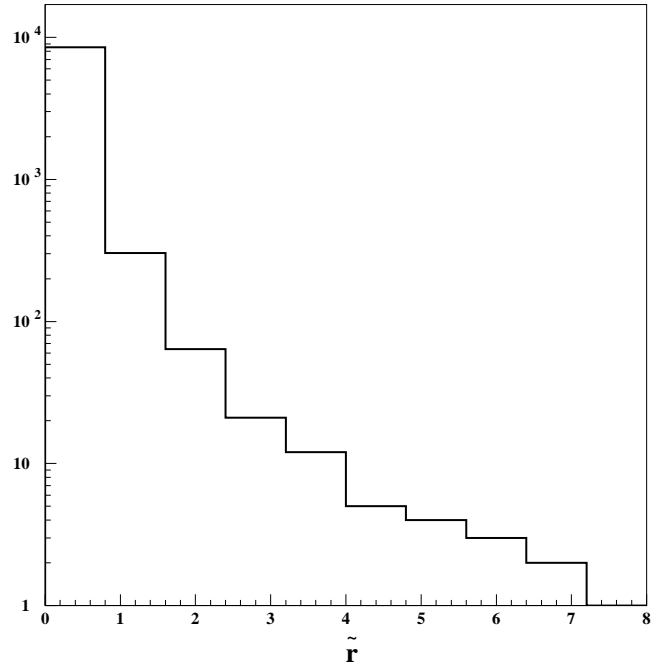}
\end{center}
\caption{\label{rtilded} Distribution of $\tilde{r}$ for microlensing events in the direction of Galactic bulge from the Monte-Carlo simulation. The average value of $\tilde{r}$ is $0.19$. Y-axis is in the logarithmic scale.
}
\end{figure}

 Taking into account
that source stars with positive planet signal is located almost at $< D_S> = 7.49$ kpc, 
this distance corresponds to the distance modulus of $\mu = 14.37$.  On the other hand from studying red clumps in the Galactic Bulge, the average extinction of stars is $<A_I> = 1.90$ \citep{pac3}. Then the average overall distance modulus with taking into accout the extinction for the direction of Bulge events would be $\mu = 16.27$. We use Table I from 
\cite{oglebinary} and obtain the absolute magnitude of stars in I-band ranges from $-0.64$ to $3.00$.  Comparing with the Hipparcos absolute magnitude \citep{hipp}, we can conclude that these stars 
are in the bright part of stellar population. In the gravitational microlensing experiment, we have a bias for  observing  bright stars as the source of microlensing events in direction of Galactic bulge.  Using the habitable zone for the orbital radius of source stars from equation (\ref{rtidle})
, we plot in Figure ({\ref{rtilded}) the distribution of $\tilde{r}$ where the microlensing events have been triggered with the microlensing surveys. 
The average value of this parameter is obtained as $<\tilde{r}> =0.19$.

Now we compare the size of $\tilde{r}$ with the distance between the binary lenses (i.e. $d$). The average value of  $d$ from the observations of $15$ binary lenses in OGLE 2002-2003 season is $<d> = 1.26$. 
The comparison of these two length scales provides the probability of crossing of planet's orbit with the 
caustic lines of binary lenses. We note that due to binary feature of microlensing light curve from the source star, source star crosses the caustic lines. Also for the case of equal mass binary lenses, in the intermediate regime where the distance of lenses is in the range of $1/\sqrt{2}<d<2$, the caustic lines are connected bar shape on the lens plane \citep{dominik99} where comparing the intermediate regime for the caustic structure with Figure (\ref{d}) means 
that almost all the cases of binary lenses are in this regime. 


In order to estimate the probability of caustic crossing of a planet, we take a simple faced-on
orbital configuration of planet around the source star and consider the relative motion of the projection 
of the planet's orbit on the lens plane with respect to the binary lens. Since $\tilde{r}\ll d$, the probability 
of caustic crossing of planet can be estimated by 
\begin{equation}
P \simeq \frac{d - \tilde{r}}{d},
\end{equation}
where using the numerical values for $\tilde{r}$ and $d$ results in $P\simeq 0.85$. 
Taking into account that $1.75 \%$ of microlensing events in our sample are binary 
lenses, then the probability of caustic crossing of a habitable planet in the binary lenses would be $1.5 \%$ of 
all the microlensing events.



All our calculation is based on geometric optics, however having a point mass source with the longer wavelengths emission from the source produces wave optics features from the microlensing magnification and result is suppressing the magnification and avoiding the singularities. In the appendix (\ref{A}) detailed calculation on the wave optics and its effect on the result of this section is discussed. We show that even using the gravitational lensing in the wave optics formalism doesn't change the results of this section.

\section{Statistics of Intelligent life detection in follow-up Microlensing observations}
\label{statint}
We have seen in the previous section that the number of planet illumination from the the single lensing is almost one order of magnitude more than that in the binary lensing. However, from the the analysis of Monte-Carlo simulation, for almost $85$ percent of the binary lensing there will be caustic crossing of planet, means that there will be high magnification by illuminating signals from the planet. Using binary lensing events for the follow-up observations has two practical advantages of, (a) in the binary microlensing events, from the deviation of light curve, we can identify the binary microlensing systems in advance where the probability of illumination of habitable planet is almost $85$ percent, (b) since the probability of caustic crossing of planet is high, SKA telescope can be used in target of opportunity mode for the follow-up observation.

In what follows we estimate of number of intelligent communication life in our observation method. We use the Drake equation which provides the probability of existing a Communicating Intelligent Life (CIL) in our Galaxy. This equation is given by 
\begin{equation} 
N_{CIL} = N_S f_P n_E f_L f_I f_C L/L_S,
\end{equation}
where $N_S$ is the number of stars in our galaxy, $f_P$ is the fraction of stars with planet, $n_E$ number of Earth like planet in the habitable zone, $f_L$ is the fraction of planets that life is developed, $f_I$ is the fraction of those planets that evolve intelligent life and $f_C$ is the fraction of those intelligent life that can have radio communication. $L$ is the time scale that intelligent civilization exists and $L_S$ is the overall age of star in the stable position.


There are optimistic and pessimistic estimations for different factors in Drake equation. We adapt updated estimations for each element. Observations of extrasolar planets with microlensing  reveals that stars can have more than $1.6$ Earth like planet means that $f_p$ is at least one \citep{cassan}. On the other hand recent analyses of the Kepler satellite showed that about $20\%$ of all Sun-like stars have Earth-sized planets orbiting within the habitable zone \citep{pet} (i.e. $n_E = 0.2$). Also HARPS team estimated that almost $30\%$ of Solar type stars have planet with an Earth mass or Super earth mass \citep{harps}. If we assume that in these planets life eventually develop, means that the probability of formation of life is one, then we can write $f_p n_E f_L \simeq 0.2$.

The later factors of $f_I$,  $f_C$ and  $L$ is unclear since only we have data from one case that exists on the Earth. We don't know what fraction of life supporting planets eventually develop to an intelligent life. Does it a fixed point of life evolution and with the probability of unity we will have an intelligent life or the probability of intelligent life is rare. Does emerging intelligent life need a long time in the orders of billions of years or it can happen in a short time scale? Also how long an intelligent life can exist. Would it disappear by self destruction or it may long live with spreading over other planets, even in the form of artificial intelligent life. We recall the multiplications of these factors by $R = f_I f_C L/L_S$.


Now we can rewrite the Drake equation for the observation of a CIL with microlensing and apply the numerical values for some the Drake factors. The result is
\begin{equation}
N_{CIL,\mu} \simeq R \times N_\mu,  
\label{nc}
\end{equation}
where subscript $\mu$ represent for microlensing events. For the parameter of $R$ we take an optimistic value assuming that every planet that supports life will eventually develop an intelligent life (i.e. $f_I = 1$) and every intelligent life eventually develops an intelligent civilization (i.e. $f_C = 1$). For the life time of an intelligent civilization (i.e. $L$), while no signal from an intelligent civilization is detected yet, however we can assume that the present 
epoch of cosmology (after finishing the star burst epoch and present calm situation), there might be a burst of intelligent civilization in the Universe. So let us take $L$ in the order of hundred million years, while there are pessimistic estimation 
of the order of $200$ years \citep{estimation}.  The hypothetical longer time scale for the life time of intelligent civilization could be due to developing and spreading of artificial intelligent life at larger areas of Galaxy.  Then let us take $L/L_S$ in the order of $10^{-2}$. Substituting numerical values for all the prameters, the number of intelligent life detection with SKA follow up microlensing observation would be $N_{CIL,\mu} \simeq 10^{-2} \times N_\mu$.  
 
We can estimate  $N_{CIL,\mu}$ events for two cases of single and binary lensing. Assuming $2000$ microlensing observations per year, we would expect  to have $N_\mu = 30$ illuminating events from the binary and $N_\mu = 50$ illuminating events from the single lensing where for the binary lensing $N_{CIL,\mu}$  is in the order of $0.3$ event per year and for the single lensing, it is in the order of $0.5$ event per year. The next generation of microlensing events will certainly increase the number of microlensing candidates per year by probing larger areas as including the spiral arm directions and lowering the limiting magnitude of observation. So we may reach in near future to the number of one microlensing event with possible intelligent life signal from the binary lensing events (taking into account optimistic 
values for the parameters of Drake equation).

As we noted before while the probability of detection of Intelligent civilization with the binary lensing is less than that of single lensing, using the early warning system, we can aware of caustic crossing of planet and as we discussed before, the probability of this event is about $85\%$. 
This makes the possibility of using the strategy of target of opportunity in the observation of intelligent life with the SKA telescope

\section{Conclusion}
\label{conc}
In this work, we proposed using the gravitational microlensing as a natural telescope for magnifying signals from an intelligent life at the Galactic scale. Microlensing can illuminate signals of a planet by single or binary lensing mode for a short period of time and if in this period we observe these target with a radio telescope, these signals can be detected. We assumed that intelligent life is using the same wavelengths that we are using in our telecommunications. Telescopes as Square Kilometre Telescope (SKA) which is designed for the cosmological observation is suitable for the observations of these signals. 

We assumed an Earth-like military transmitter on a planet with intelligent civilization and took SKA telescope for 
detecting transmitted signals. Then we simulate the number of microlensing events with enough illuminations from the planets orbiting within the habitable zone that can be detected by the SKA telescope. In order to estimate the number of events, we performed a Monte-Carlo simulation and obtained the number of high magnification events in both single and binary lensing 
channels. We have shown that taking into account the wave optics effects in the gravitational lensing doesn't change  results of geometric optics. The result of simulation is that for single-lens microlensing events for $2.5\%$ of overall event we will observed illumination of planet around the source star and for the binary lensing that is about $1.5\%$ of all the events.


Finally we used Drake equation in our study and adapted the factors of this equation from the recent observations to estimate the number of intelligent life detection. For the factor of probability of intelligent life formation (i.e. $f_I$) and communicating civilization (i.e. $f_C$), we adapt the optimistic values. Also we assumed that intelligent life might be developed in the form of artificial intelligent civilization and spread in the local areas of Galaxy. The annual number of events, performing follow-up observation of microlensing events with SKA telescope and getting a positive signal is about $0.5$ for illuminating events of single lensing and $0.3$ for the case of binary lensing. While the number of binary lensing events is less than single lensing, the probability of illuminating effect for the binary lensing is almost $85\%$. One of the advantages of binary microlensing events is that it can be identified long before the caustic crossing, using early warning system of surveys. 
On the other hand, having binary lensing events, it can be observed by SKA as target of opportunity mode which make this program feasible. Increasing the number of microlensing events in the direction of Galactic bulge as well as spiral arms of Galaxy is achievable with the next generation of microlensing surveys, using wider field of view and increasing the sensitivity of detectors. This method of SETI observations would be a new item for the astrophysical applications of gravitational microlensing observations \citep{rahvar}.

\acknowledgments 
We would like to thank anonymous referee for his/her useful comments that helped for improving this work.

\appendix
\section{Wave optics effect and suppressing of signals}
\label{A}
A point like astronomical object produces a flat wave front at large distance from the source. Let us assume that we put a Young double slit experiment in front of this wave front.  Slits at suitable separation results in fringes on the observer plane. The gravitational lensing system can play the same role of Young double slit experiment by producing double or multiple images on the lens plane . The result would be interference on the observer plane where slits are locates at the astronomical scales far from the observer. The wave optics property of light in gravitational lensing is extensively discussed in \cite{Schneider}. This method also is a complimentary observation for detection of planets around the lenses \citep{ms}.

In geometric optics, the caustic lines on the source plane are singular lines where the Jacobian of determinant between the source plane and image plane is zero. However in the wave optics since the intensity of light is distributed in the diffraction fringes, the singularity is resolved. The magnification factor from a point like source near the caustic line is given by 
\begin{equation}
\mu(y) = \frac{\mu_{max}}{Q^2}\left[Ai(\frac{y}{Y_0})\right]^2,
\end{equation}
where $y$ is the distance of source from the caustic line, $Ai(x)$
is the Airy function and $Q\simeq 0.5357$ is the maximum value of
Airy function. The maximum magnification of the fringes as well as
the wavelength of Airy function depends on the Fermat potential as
\begin{equation}
\mu_{max} = \frac{2^{5/3}\pi
f^{1/3}}{|\phi_{11}||\phi_{222}|^{2/3}}Q^2,~~~~~ Y_0 =
\left(\frac{|\phi_{222}|}{2f^2}\right)^{1/3}
\label{mmax}
\end{equation}
where $\phi$ is the Fermat potential for a binary lens at the
position of the images, subscript is the spatial derivative in two
directions and $f$ is the multiplication of wavenumber to the
Schwarzschild radius of lens $f = 2k R_s$. For a binary lenses the
Fermat potential is given by
\begin{equation}
\phi(\mathbf{x},\mathbf{y}) = \frac{1}{2}(\mathbf{x}-\mathbf{y})^2 -
q_1\ln(|\mathbf{x}-\mathbf{x_1}|) -
q_2\ln(|\mathbf{x}-\mathbf{x_2}|),
\end{equation}
where $q_1$ and $q_2$ are the relative mass of lenses to the total
mass of system. $\mathbf{x}$ is the position of image normalized to
the overall Einstein radius, $\mathbf{y}$ is the position of source
normalized to the projected Einstein radius of binary system and
$\mathbf{x_1}$ and $\mathbf{x_2}$ are the position of lenses. The lens equation from the 
Fermat potential obtain by extremum condition of this function (i.e. $\partial \phi(\mathbf{x},\mathbf{y})/\partial\mathbf{x} = 0$). 

Detailed calculation of Fermat potential provides $|\phi_{222}|^{-2/3} |\phi_{11}|^{-1} \simeq q^{1/3}$ and the denominator of equation (\ref{mmax}) for equal mass lens is in the order of $0.8$. For nominator of equation (\ref{mmax}), we adapt one meter size electromagnetic wavelength for intelligent life transmitter, also one solar mass for the mass of binary lenses. The result is the maximum magnification of planet's signal, $\mu_{max} = 75$, during the caustic crossing. Using equation (\ref{minimum}) for minimum amount magnification for the detection, we obtain the limiting distance of $D_S<10$ kpc for observability, which is well within the observable distance of sources in Bulge microlensing events.  Hence the wave optics formalism doesn't change the results of what is calculated based on gravitational lensing in the geometric optics.

\end{document}